\documentclass[letter]{aa}  

\usepackage{graphicx}
\usepackage{txfonts}
\usepackage{lipsum}
\usepackage{subcaption}         
\usepackage{lscape}             
\usepackage{placeins}           

\begin{document}

   \title{Discovery of a persistent radio source associated with FRB\,20240114A}


   \author{G. Bruni\inst{1}
        \and L. Piro\inst{1}
        \and Y.-P. Yang\inst{2,3}
        \and E. Palazzi\inst{4} 
        \and L. Nicastro\inst{4}
        \and A. Rossi\inst{4}
        \and S. Savaglio\inst{5,4,6}
        \and E. Maiorano\inst{4}
        \and B. Zhang\inst{7,8}
        }

   \institute{INAF -- Istituto di Astrofisica e Planetologia Spaziali, via Fosso del Cavaliere 100, 00133 Rome, Italy\\
             \email{gabriele.bruni@inaf.it}
             \and
            South-Western Institute for Astronomy Research, Yunnan University, Kunming, China. 
            \and
            Purple Mountain Observatory, Chinese Academy of Sciences, Nanjing, China.         
             \and
            INAF -- Osservatorio di Astrofisica e Scienza dello Spazio di Bologna, via Piero Gobetti 93/3, I-40129 Bologna, Italy
            \and
            Dipartimento di Fisica, Università della Calabria, Arcavacata di Rende, Italy 
            \and
            Laboratori Nazionali di Frascati, INFN (Istituto Nazionale di Fisica Nucleare), Frascati, Italy 
            \and
            Nevada Center for Astrophysics, University of Nevada, Las Vegas, NV, USA. 
            \and
            Department of Physics and Astronomy, University of Nevada, Las Vegas, NV, USA
            }

   \date{Received November 30, 20XX}


  \abstract
   {}  
   {We present the discovery of a fourth persistent radio source (PRS) associated with a fast radio burst (FRB).} 
   {Following previous indications of a candidate PRS associated with FRB\,20240114A, we performed deep Very Long Baseline Array observations at 5 GHz to test the presence of a compact radio source within the uncertainty position of this FRB ($\pm$200 mas).}
   {We detected a component $\sim$50 mas northwards of the nominal position provided by the PRECISE collaboration. The corresponding radio luminosity together with the Faraday rotation measure provided by previous observations of the FRB locate this PRS in the expected region of the radio luminosity versus Faraday rotation measure relation for the nebular model, further supporting its validity. Comparison of the measured flux density with the values collected at a lower frequency by previous studies indicates a possible steepening of the radio spectrum in the 1--5 GHz range. Optical observations performed with the Large Binocular Telescope could reveal that the FRB and its PRS lie at $\sim$1 kpc from the centre of the host galaxy, which is a dwarf sub-solar metallicity starburst galaxy with a star-formation rate of $\sim 1 M_\odot\;\mathrm{yr^{-1}}$ and a stellar mass of $M\sim10^8 M_\odot$.}
   {}


   \maketitle

\section{Introduction}
Fast radio bursts (FRBs) are bright millisecond-duration sources of extragalactic origin, and so far they  have only been observed at radio wavelengths. The vast majority ($>700$) are one-off events, but several are repeating -- referred to as repeating FRBs (rFRBs; see e.g. \citealt{2023RvMP...95c5005Z}). It is not unlikely that the repeating nature is ubiquitous but often too faint to be captured \citep{Kirsten+24}. Notwithstanding the large number of events and the substantial efforts to carry out deep multiwavelength observations, the lack of counterparts at other wavelengths (with one possible exception of a magnetar in our Galaxy \citep{Bochenek20}) has left the door open for various progenitor models.

Magnetars can reproduce many of the observed properties \citep{Zhang20} and are one of the leading scenarios for FRB central engines. However, even this kind of object can be formed by different progenitor channels, for instance, in young (through core-collapse supernovae) or older (through compact binary coalescence) stellar environments \citep{Margalit19,niu22}, as seen for FRB\,20200120E in a globular cluster \citep{Kirsten22}. Such different progenitors can in principle account for different properties of the host and local environment, but the present sample is still too quantitatively and qualitatively limited to settle the issue. 
An important step forward has been pioneered in the past years by extending the study of the environment from radio to X-rays, including high-resolution optical and IR spectrophotometry, down to the sub-arcsecond level  \citep{2021A&A...656L..15P,Bruni2024Nat}. This approach  has proven to be highly rewarding, as it allows for characterisation of the host galaxies and proper mapping of their star formation regions (e.g. \citealt{2017ApJ...834L...7T,Bhandari20,Bhandari22}) and singles out the presence of a low luminosity and continuous radio source, or persistent radio source (PRS), connected to the central engine in FRB\,20121102A \citep{chatterjee17,marcote17}, FRB 20190520B \citep{niu22}, and FRB\,20201124A \citep{Bruni2024Nat}. 


\begin{figure*}
    \centering
    \begin{tabular}{cc}
      \includegraphics[width=0.43\linewidth]{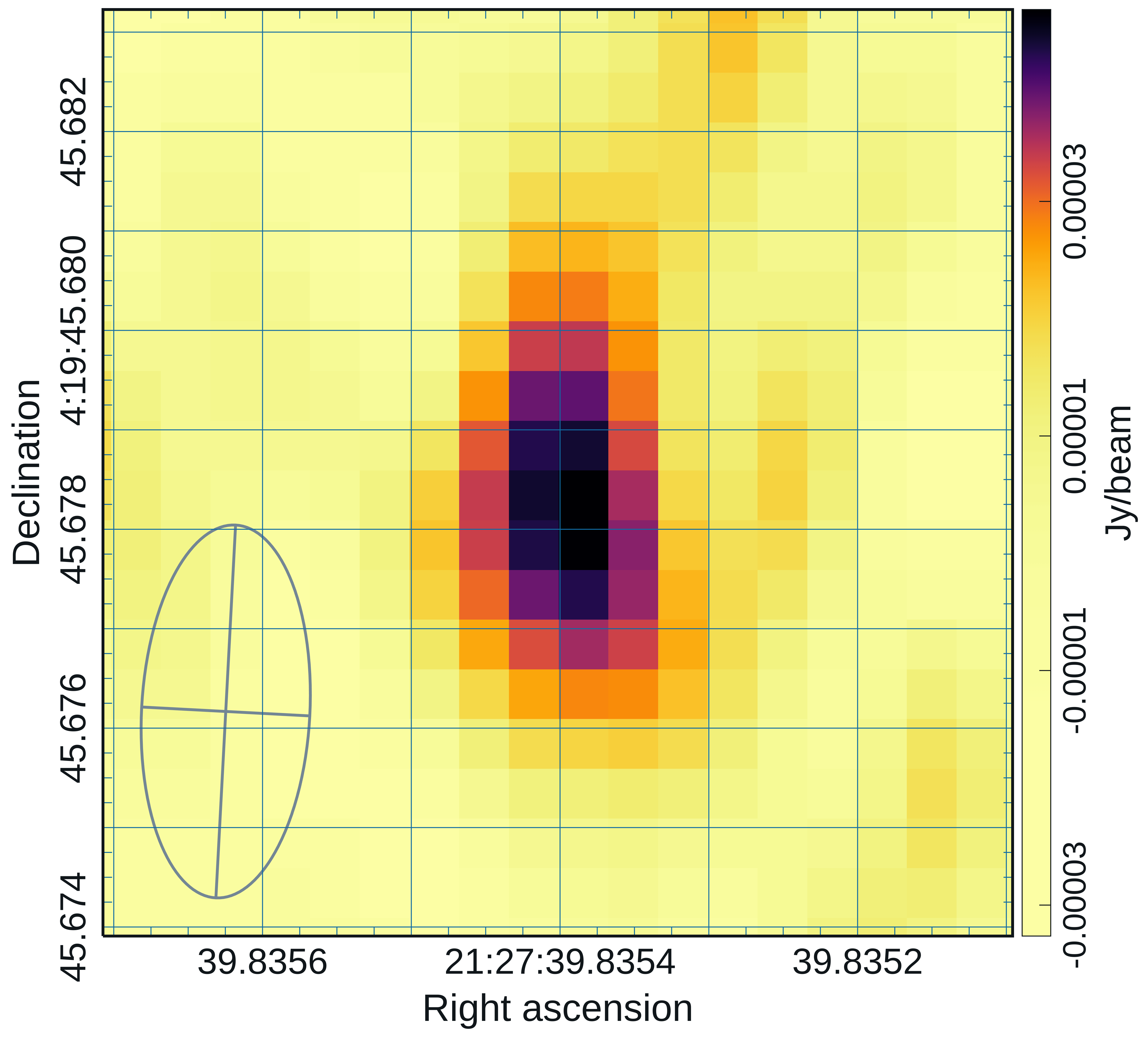} &
    \includegraphics[width=0.55\linewidth]{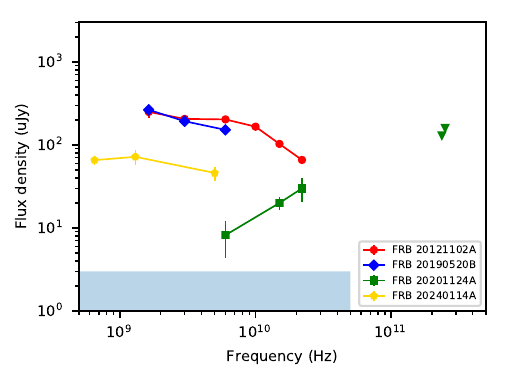}  \\
    \end{tabular}
    
    \caption{Left panel: Very Long Baseline Array image of the PRS at 5 GHz. The FWHM (3.7$\times$1.7 mas) is shown in the lower-left corner. Right panel: Radio spectral shapes of the four PRSs discovered so far. This image was built with VLA data (except for our VLBA 5 GHz measurement). The blue shaded area represents the region below the 3-$\sigma$ VLA sensitivity in the range 1–50 GHz.}
    \label{fig:PRS+sed}
\end{figure*}


The discovery of compact PRSs associated with three rFRBs has represented important progress in the study of possible FRB progenitors. Such compact emission may hold the key to unveiling the central engine. Highly active repeaters could be newborn magnetars still residing in their complex magnetised natal environment. The large Faraday rotation measure |RM| values of some FRBs imply a dense and magnetised environment close to the sources, which led to the idea of a simple relation between the luminosity of the PRS and the |RM| that traces the magneto-ionic medium in the FRB environment \citep{yang20,yang22}. If a population of relativistic electrons is maintained within such a magnetised medium, it will naturally produce synchrotron radiation resulting in a bright PRS \citep{Murase16, margalit18}. The recent discovery of the PRS associated with FRB\,20201124A has validated the relation by extending the explored range of RM and luminosity by two orders of magnitude, and it supports the idea that the persistent radio emission is of nebular origin \citep{Bruni2024Nat}. Moreover, a Very Large Array (VLA) systematic search towards 37 CHIME repeating FRBs revealed two additional candidate PRSs that fall in the expected region of the luminosity versus |RM| relation \citep{Ibik2024}. Since nebulae can originate from magnetars and hyperaccreting X-ray binaries, these findings are crucial to disentangling different scenarios and eventually circumscribing the nature of the FRB central engine. 

\begin{table*}[]
    \centering
    \caption{Journal of radio observations.}
    \begin{tabular}{cccccccccccc}
    \hline
    Telescope   &   Date         &  Frequency   & FWHM          &  RMS      &   Flux density    \\ 
                &   (dd/mm/yyyy) & (GHz)        & (mas)         & ($\mu$Jy/beam) &   ($\mu$Jy)  \\
    \hline
    e-Merlin    &  20--28/09/2024 & 5            & $228\times 87$ &  10        &   $<50$           \\
    VLBA        &  28,30/09/2024 & 5            & $3.7\times 1.7$    &   8        &   $46\pm9$        \\
    \hline
    \end{tabular}
    \label{tab:radio}
\end{table*}

\section{The hyperactive FRB\,20240114A}

Recently, a further nearby and very active rFRB was found by CHIME/FRB (FRB\,20240114A, \citealt{2024ATel16420....1S}) and associated with a nearby galaxy \citep[$z=0.13$;][]{Bhardwaj2024ATel}. The localisation provided by the PRECISE collaboration through EVN observations of the bursts \citep{2024ATel16542....1S} allowed the uncertainty to be lowered to $\sim$200 mas. 
Since its discovery, FRB\,20240114A has been very active, showing burst rates up to $\sim 500$ bursts per hour in the 1.0--1.5 GHz band \citep{Zhang2024ATel}. 

Further radio continuum observations at a few arcseconds resolution allowed for the identification of a candidate PRS, first through a reanalysis of MeerKAT data \citep{2024ATel16695....1Z,2025arXiv250114247Z} that detected a continuum emission of $72\pm14$ $\mu$Jy at 1.3 GHz and more recently through uGMRT observations at 650 MHz that detected a continuum source with a flux density of $65.6\pm8.1$ $\mu$Jy, suggesting a flat spectrum \citep{2024ATel16820....1B,2024arXiv241213121B}. Nevertheless, in-band spectral analysis of MeerKAT data resulted in a spectral index of $-1.1 \pm 0.8$ between 1 and 1.5 GHz \citep{2024ATel16695....1Z}, suggesting a possible steepening of the radio emission at higher frequencies. However, these observations at a resolution of a few arcseconds did not allow for secure association with the FRB itself.
%


\section{Confirmation of the associated PRS}

During September 2024, we performed e-Merlin and Very Long Baseline Array (VLBA) follow-up observations pointed at the FRB coordinates provided by the PRECISE collaboration. Details for the VLBA observations are discussed below. The e-Merlin observations resulted in an upper limit and are reported in Appendix \ref{app:radio}. The main quantities are reported in Table \ref{tab:radio}. 
\subsection{VLBA detection of a compact radio source}
We observed the FRB location with the VLBA at 5 GHz (C-band) under DDT project BB468 (PI Bruni). Observations were divided into two identical runs occurring on September 28 and 30, 2024, and had a total of 12 hours. Phase referencing was applied using J2123+0535 as calibrator. Data were reduced through the standard {\tt{AIPS}} procedure for continuum, and then calibrated visibilities from the two runs were concatenated and imaged in {\tt{CASA}}, achieving an angular resolution of 3.7$\times$1.7 milli-arcsec in natural weighting and an RMS of 8 $\mu$Jy/beam.

An unresolved source was identified within the $\pm$200 mas FRB positional uncertainty provided by the PRECISE collaboration $\sim$50 mas northwards of the phase centre (see Fig. \ref{fig:PRS+sed}, left panel). The peak position is RA: 21:27:39.83538, Dec: 04:19:45.6783 (J2000). Image registration was provided through the phase calibrator, whose position is known with an accuracy of 0.03 mas.\footnote{See the VLBA calibrators reference page: \url{https://obs.vlba.nrao.edu/cst/calibsource/16616}} We thus considered a standard 10\% of the full width half maximum (FWHM) major axis as a conservative estimate of the positional uncertainty (i.e. $\pm$0.4 mas). The measured flux density (peak) is $46\pm9$ $\mu$Jy, where the uncertainty is the squared sum of the image RMS plus a typical 10\% uncertainty on the absolute flux scale. The ratio between the peak and the RMS corresponds to a 5.7-$\sigma$ detection. The flux falls below the e-Merlin upper limit at the same frequency. No other peak with a significance above 5-$\sigma$ was found in the PRECISE uncertainty region. At the redshift of the host galaxy, 
the physical size of the PRS is constrained to $\lesssim 4$ pc. This resulted in a brightness temperature of $T_b>7.8\times10^{5}$ K and a luminosity of $L_{\rm radio}=2.2\times10^{28}$ $\rm erg~s^{-1}~Hz^{-1}$, placing the PRS more than one order of magnitude above the luminosity of the brightest star forming regions ($\sim10^{26}$ $\rm erg~s^{-1}~Hz^{-1}$, \citealt{Kennicutt1988}) and suggesting a non-thermal (synchrotron) origin for its radio emission. This outcome is consistent with the other PRSs studied so far.
This detection confirms the presence of a compact continuum source associated with FRB\,20240114A, making this the fourth PRS known to date -- following FRB\,20121102A \citep{chatterjee17}, FRB\,20190520B \citep{niu22}, and FRB\,20201124A \citep{Bruni2024Nat}.
%


\subsection{Radio spectrum of the PRS}
In Fig. \ref{fig:PRS+sed}, right panel, we show the radio spectral shape of each of the four PRSs discovered so far. As discussed in the literature, while the PRS associated with FRB\,20121102A and FRB\,20190520B showed a negative spectral index, the more recent one found for FRB\,20201124A showed a positive inverted spectral index. The mentioned uGMRT and MeerKAT detections for the PRS of  FRB\,20240114A suggest a flat or inverted spectrum between 0.65 and 1.3 GHz ($\alpha=0.13\pm0.33$), while the in-band spectrum in MeerKAT data (1--1.5 GHz) indicate a steepening of the spectrum ($\alpha=-1.1\pm0.8$). On the other hand, our VLBA detection allows for an estimate spectral index of $\alpha=-0.34\pm0.21$ between 1.3 and 5 GHz, but its uncertainty does not yet allow us to conclusively claim a flat ($\alpha>-0.5$) or steep ($\alpha<-0.5$) spectrum. 

\begin{figure}
    \centering
    \includegraphics[width=0.95\linewidth]{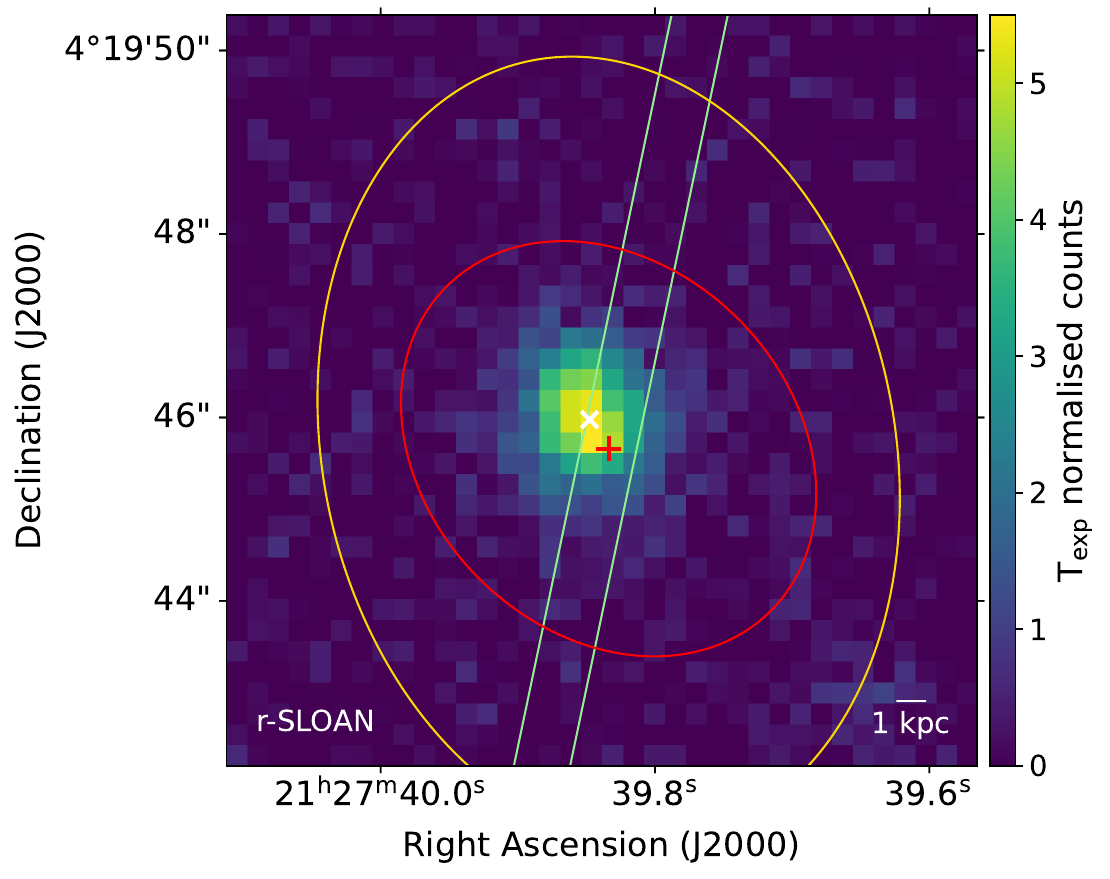}
    \caption{Large Binocular Cameras r$^\prime$ filter image. The PRS position is marked by a plus sign, and the galaxy centre is marked with an 'x'. The slit used for the MODS spectrum is shown. The image pixel scale is 0\farcs23, and the RMS is 0.25 cts. The galaxy extension is $\simeq 3\farcs2\times 2\farcs8$. The uGMRT (red) and MeerKAT (gold) beams are also shown.}
  \label{fig:LBT}
\end{figure}

The uGMRT and MeerKAT observations at a resolution of a few arcseconds encompass the whole host galaxy (see Fig. \ref{fig:LBT}): this prevents one from disentangling its possible contribution from the emission of the compact PRS. Nevertheless, when considering the standard relation between $L_{1.4\rm{GHz}}$ and star-formation rate (SFR) from \cite{murphy11}, the MeerKAT flux density should correspond to an SFR of $\sim$1.8 M$_\odot$/yr, which is much larger than our estimate based on optical spectral energy distribution (SED) fitting ($<1$ M$_\odot$/yr, see Section 4.2). Thus, the contribution of star-forming regions to the measured flux densities at a resolution of a few arcseconds should be limited. 

Further, VLBI observations in the 1--2 GHz range should pinpoint the compact PRS emission and correctly estimate its flux density. This would allow for a comparison with the MeerKAT flux at the same frequency and assessment of the spectral curvature of the PRS. If the flux density result is consistent with MeerKAT, this would confirm the spectral curvature suggested by the lower flux density at 650 MHz measured by uGMRT, resulting in the first PRS with a peaked synchrotron spectrum. This would make this PRS an intermediate case between the two steep PRSs associated with previous FRBs and the more recent inverted one associated with FRB\,20201124A.
%


\section{Host galaxy properties}
\label{sect:host}

\subsection{Optical observations and SED fitting}
Photometric observations of FRB\,20240114A were obtained in the $g^\prime r^\prime i^\prime z^\prime$ optical bands with the Large Binocular Cameras \citep[LBCs;][]{Giallongo2008a} mounted on the Large Binocular Telescope (LBT) on Mt. Graham, Arizona, USA. Data were collected on November 9, 2024, and additional data in only the $r^\prime i^\prime$ bands were acquired on the following night. The data reduction is described in Appendix \ref{app:opt}.
%
%
We estimate the galaxy centre to be RA=21:27:39.849, Dec=+04:19:46.00 (J2000) with an extension of $\simeq 3\farcs2\times 2\farcs8$ (Fig. \ref{fig:LBT}), corresponding to a physical size of $\simeq 7.7\times 6.7$~kpc for a scale of {2.4 kpc/\arcsec} at redshift $z = 0.13056$ (see Sec. \ref{speobs}). The PRS is located 0\farcs43 from the centre of the host galaxy, which corresponds to $\simeq 1$ kpc.

After correction for the Galactic foreground extinction of $\mathrm{E(B-V)}=0.06$ mags \citep{sf2011}, we modelled the SED of the host galaxy using the \texttt{Cigale}
code \citep{Boquien2019}.
We added the $u^\prime=23.0\pm0.4$ detection from the SDSS survey to our dataset (see Fig. \ref{fig:sedhost}).
In addition, we fixed the redshift to 0.13 and used a derived stellar dust attenuation derived from the one measured with the Balmer decrement ($\rm H\alpha/H\beta$; see Appendix \ref{app:opt}), 
$\mathrm{E(B-V)_{star}}=0.4 \times \mathrm{E(B-V)_{gas}}$ \citep{Calzetti1997, Asari2007}.
The galaxy model gives $\chi^2/{\rm dof}=0.25$
for a stellar mass of $M_\star = 10^{8.1 \pm 0.2}\, M_\odot$
and $\rm{SFR} < 1 \, {\it M}_{\odot}\,{\rm yr}^{-1}$. While the derived stellar mass is reliable, the SFR is degenerate with age (0.3--1.2 Gyr).
%


\subsection{Spectroscopic observations and data analysis}
\label{speobs}
On November 10, 2024, a set of three spectroscopic exposures of 1200 s each were performed with the Multi-Object Double Spectrograph MODS-1 \citep{Pogge2010a} mounted on LBT. We used the dual-grating mode, which provided a wavelength coverage of 3200--9500~\AA, and a slit mask with a width of $1\farcs2$. The slit was intentionally placed with a position angle of $-11\fdg 9$ and centred on the position of the PRS, and therefore it did not cover the whole galaxy (see Fig. \ref{fig:LBT}).
A redshift of $z = 0.13056 \pm 0.00003 $ was determined (see Appendix \ref{app:opt} for the details).
%
%
By using the dust-corrected fluxes and the prescription provided by \cite{Kennicutt1988} and \cite{Savaglio2009}, we derived $ \rm{SFR(H}\alpha) = 0.36\pm 0.01$ M$_\odot$/yr. We note that this is the SFR measured in the region of the galaxy covered by the slit, which is 40\% of the whole galaxy extension. Consequently, the total SFR is likely 
higher than the measured value, $ \rm{SFR(H}\alpha) \simeq 0.9$. The resulting specific SFR (sSFR; the SFR/$M_\star$) for the entire galaxy is $-8.5 < \log (\mathrm{sSFR}) < -8.1$ yr$^{-1}$, which is one of the highest sSFR values for an FRB host (see e.g. \citealt{Nicastro2021}), and it is typical of a galaxy in a `starburst' regime (but see also \citealt{chen2025}). 

We also used the dust corrected line fluxes (see Tab. \ref{tab:spelines}) to compute the following ratios:  [S\,$\rm II]/H\alpha = -0.910 \pm 0.007$, [N\,$\rm II]/H\alpha = -0.998 \pm 0.003$, and [O\,$\rm III]/H\beta = 0.498 \pm 0.004$, which places the host galaxy within the star-forming region of the Baldwin-Phillips-Terlevich (BPT; \cite{BPT1981}) diagram
%
(see Figs. \ref{fig:BPT_NII}, \ref{fig:BPT_SII}).
%
To estimate the gas metallicity, we used the method proposed by 
\cite{Curti2017}, which gives an average metallicity from all possible metallicity calibrators. This yielded $12+\log(\textnormal{O/H})\sim 8.5$. This metallicity, although sub-solar, is relatively high if one considers the mass-metallicity relation for galaxies with a similar stellar mass at $z\sim0.1$ \citep{Bassini2024}.
%
\begin{figure*}[t]
    \centering
    \includegraphics[width=0.8\linewidth]{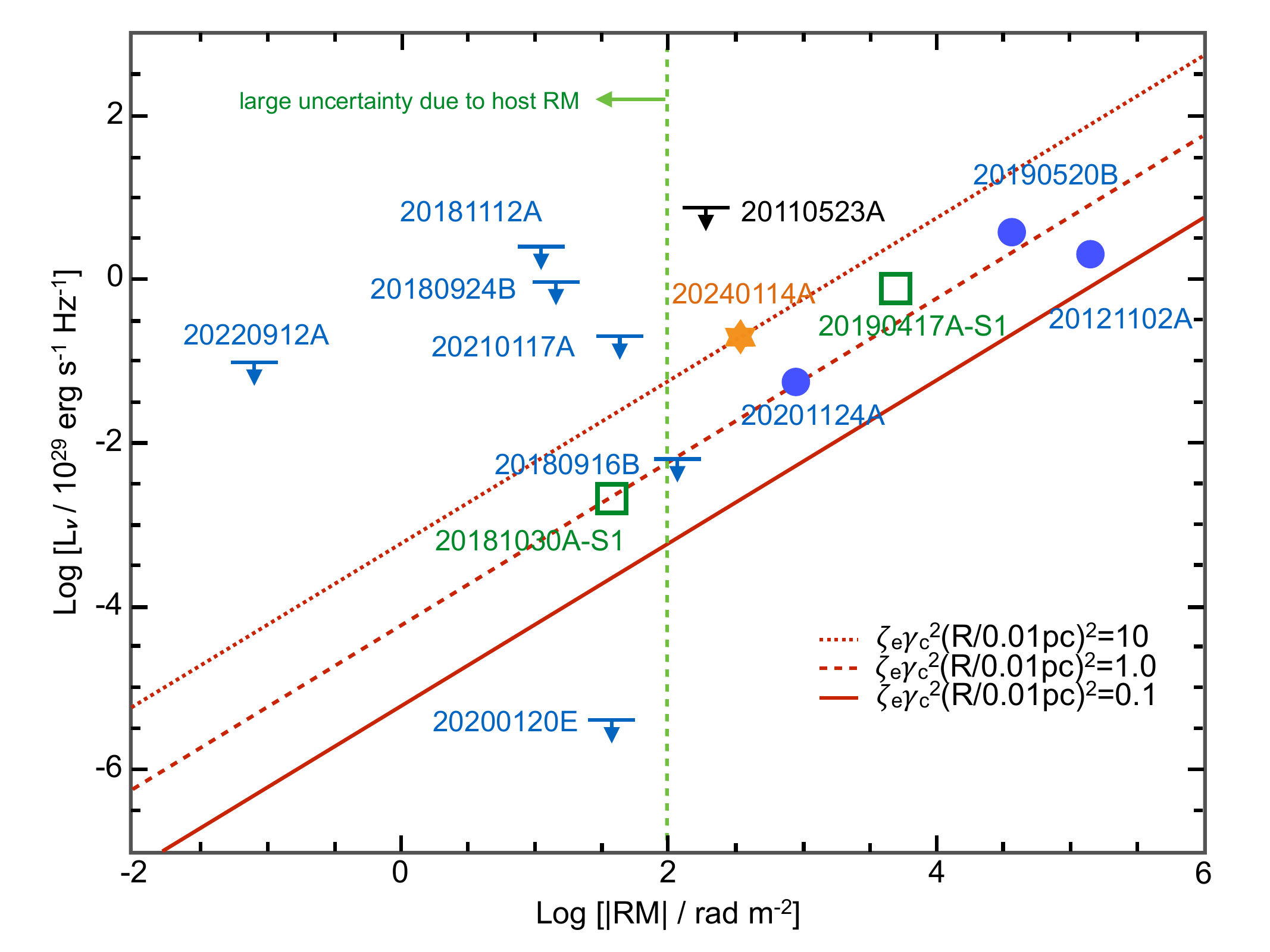}
    \caption{Proposed relation between the PRS specific radio luminosity and the FRB RM adopting the general nebula model from \cite{yang20, yang22}
    The red dotted, dashed, and solid lines denote the predicted relations for $\xi_e\gamma_c^2(R/0.01~{\rm pc})^2=10,1.0$, and $0.1$, respectively. The blue circles denote the FRBs with measured persistent emission flux and RM (FRBs 20121102A, 20190520B, and 20201124A). The green squares denote the PRS candidates of 20181030A-S1 and 20190417A-S1 given by \cite{Ibik2024}. The black upper limit indicates the non-localised FRB\,20110523A with an upper limit of the persistent emission and a measured value of RM, which gives one of the most conservative constraints for non-localised sources in the literature.  The blue upper limits correspond to the FRBs with precise (arcsecond) localisations. Owing to the large RM uncertainty of the host interstellar medium, the data of FRBs with $|{\rm RM}|\lesssim100~{\rm rad~m^{-2}}$ (on the left side of the green dashed line) might substantially deviate from the predicted relation.}
    \label{fig:relation}
\end{figure*}


\section{Implications for the nebular model}

As shown in Fig. \ref{fig:PRS+sed} (right panel), the SED of the PRS associated with FRB\,20240114A is similar to those of the PRSs associated FRB\,20121102A and FRB\,20190520B. Since there is no statistically significant claim of the spectral peak's location from the present observations, the position of the spectral peak might be below the uGMRT band (i.e. $\nu_{\rm peak}\lesssim0.65~{\rm GHz}$), which could constrain the properties (e.g. magnetic field, electron density distribution, etc.) of the PRS emission region (see Appendix \ref{calculations} for more details).

\cite{yang20, yang22}
proposed that the RM of an rFRB and the associated PRS might originate from the same region. In this case, there is a simple relation between the RM and the PRS luminosity that is generic and weakly depends on the PRS model.
The PRS of FRB\,20240114A has a flux density of $F_\nu=46\; {\rm \mu Jy}$ for the VLBA observation. Using the host-galaxy redshift of $z=0.13$, the specific luminosity of the PRS is $L_\nu\simeq2.2\times10^{28}~{\rm erg~s^{-1}Hz^{-1}}$. The RM of FRB\,20240114A is measured to be ${\rm RM}\simeq338.1\; {\rm rad~m^{-2}}$ \citep{Tian2024}. In Fig. \ref{fig:relation}, we plot the relation between the specific luminosity of the PRS against the RM of known FRBs. One can see that the measurements of the FRBs 20121102A, 20190520B, 20201124A, 20240114A, and two other PRS candidates (20181030A-S1 and 20190417A-S) satisfy the predicted relation well. Meanwhile, the observed specific PRS luminosity of FRB\,20240114A is close to the value of $\zeta_e\gamma_c(R/10^{-2}\; {\rm pc})^2\sim10$ (see Appendix \ref{calculations} for more details).



\section{Conclusions}
The findings of this observational work, and their theoretical implications, are summarised below:
\begin{itemize}
    \item Through VLBA observations at parsec-scale resolution, we discovered a compact continuum source consistent with the position of FRB\,20240114A. Its flux density, brightness temperature, and radio luminosity suggest a non-thermal origin and a negative spectral index, as expected from previous lower frequency and resolution MeerKAT observations. These findings suggest that the source is a PRS associated with FRB\,20240114A.
    \item The radio luminosity and Faraday rotation measure place the PRS in the expected region of the L versus RM theoretical relation, further supporting its validity and strengthening the case for a nebular origin for the continuum emission associated with some FRBs.
    \item The host galaxy is a dwarf sub-solar metallicity starburst galaxy with spectral properties that exclude an AGN origin for the PRS. 
\end{itemize}
Further VLBI observations will better characterise the radio spectral shape of this peculiar PRS and eventually provide convincing evidence of the presence of a peak. The latter could further constrain the nebular model and give information on the plasma energetics.
\begin{acknowledgements}
The research leading to these results has received funding from the European Union’s Horizon 2020 programme under the AHEAD2020 project (grant agreement no. 871158).
Y.P.Y. is supported by the National Natural Science Foundation of China grant No.12473047, the National Key Research and Development Program of China (2024YFA1611603) and the National SKA Program of China (2022SKA0130100).
A.R. acknowledge INAF project Supporto Arizona \& Italia.
We thank the staff of the e-Merlin, in particular Dr. David Williams-Baldwin, for their prompt support and assistance with these observations. e-MERLIN is a National Facility operated by the University of Manchester at Jodrell Bank Observatory on behalf of STFC, part of UK Research and Innovation.
The National Radio Astronomy Observatory is a facility of the National Science Foundation operated under cooperative agreement by Associated Universities, Inc. This work made use of the Swinburne University of Technology software correlator, developed as part of the Australian Major National Research Facilities Programme and operated under licence.
We thank the staff of LBT Observatory and LBT-Italy, in particular D. Paris, E. Marini, and F. Cusano in obtaining these observations. The LBT is an international collaboration of the University of Arizona, Italy (INAF: Istituto Nazionale di Astrofisica), Germany (LBTB: LBT Beteiligungsgesellschaft), the Ohio State University, representing also the University of Minnesota, the University of Virginia, and the University of Notre Dame.
\end{acknowledgements}

   \bibliographystyle{bibtex/aa} 
   \bibliography{Main} 
%
\begin{appendix}
\section{e-Merlin upper limit}
\label{app:radio}

Observations with the enhanced Multi Element Remotely Linked Interferometer Network (e-Merlin) were carried out in five $\sim$12-hour runs between September 20 and 28, 2024, at 5 GHz (C-band), under project CY18016 (PI Bruni). Phase referencing was applied. Visibilities were calibrated through the e-Merlin pipeline \citep{2021ascl.soft09006M}, and imaged in {\tt{CASA}} \citep{2022PASP..134k4501C}. An RMS of 10 $\mu$Jy/beam was reached, and no source was detected either within the uncertainty region of the FRB coordinates ($\pm200$ mas), or within the wider image field ($20\arcsec\times 20\arcsec$). This sets a 5-sigma upper limit of 50 $\mu$Jy for the candidate PRS flux density at 5 GHz. The image resolution in natural weighting was $228\times87$ mas.


\section{LBT observations and data reduction}
\label{app:opt}

\subsection{MODS spectroscopy}
Optical spectroscopy of FRB\,20240114A was obtained with the 
MODS-1 mounted on LBT. Only the MODS-1 arm of the binocular telescope was available during the observations. We employed the dual-grating mode (grisms G400L and G670L) providing a wavelength coverage range 3200--9500~\AA, and a slit mask with a width of $1\farcs2$.
Observations were carried out on November 10, 2024, and consisted of three exposures of 1200 s. Slit-alignment was done after blind offsets from a $R\sim$17 mag star in the field. The slit was centred on the position of the PRS with a North-to-East position angle of $-11\fdg 9$, therefore covering only partially the galaxy (Fig. \ref{fig:LBT}).

We processed the spectral data using the Spectroscopic Interactive Pipeline and Graphical Interface (SIPGI) tool, which is specifically designed for the reduction of MODS and LUCI spectra \citep{sipgi2022}. We first applied a bad pixels map (generated using imaging flats) to each observed frame, along with a cosmic ray correction. Afterwards, each frame underwent independent bias subtraction and flat-field correction, using a master flat derived from a set of spectroscopic flats. For wavelength calibration, we used the inverse solution of the dispersion, calculated from arc lamp frames and saved in the master lamp. It was then applied to individual frames to calibrate the wavelength and correct for optical distortion. We achieved a wavelength calibration accuracy of $\sim 0.067 \AA$ for the red and $\sim 0.052 \AA$ for the blue channel. Next, we extracted two-dimensional, wavelength-calibrated spectra and performed sky subtraction. To obtain flux-calibrated spectra, we applied the sensitivity function derived from the spectro-photometric standard star Feige 110 observed immediately after the FRB. Finally, the wavelength- and flux-calibrated, sky-subtracted spectra were combined, and the one-dimensional spectrum of each source was extracted.

\begin{figure*}
    \centering
    \includegraphics[width=0.95\linewidth]{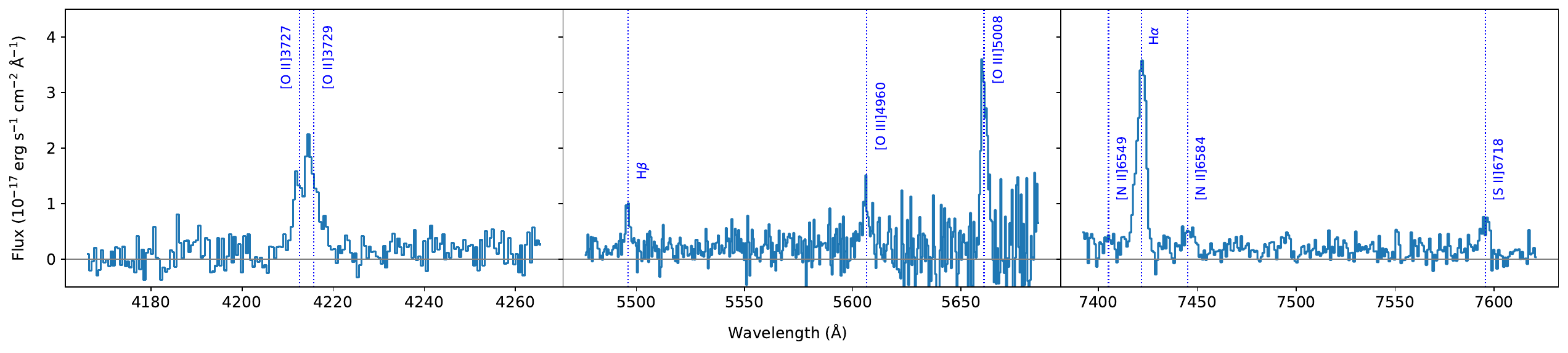}
\caption{
Optical emission lines detected with LBT/MODS at the position of the PRS.
The three panels show spectral intervals around the lines: the [O\,$\rm II$] doublet (left panel); the H$\beta$ and [O\,$\rm III$] doublet (central panel); H$\alpha$, [N\,$\rm II$]$ \, 6549$, [N\,$\rm II$]$\,6584$, and [S\,$\rm II$]$ \, 6718$ (right panel).
}
\label{fig:spelines}
\end{figure*}

Figure \ref{fig:spelines} shows sections of the spectra around the detected emission lines whose fluxes are reported in Tab. \ref{tab:spelines}.
A redshift of $z = 0.13056 \pm 0.00003$
was determined from the simultaneous detection 
of the following emission lines: H$\alpha$, H$\beta$, [N\,$\rm II$]$ \, 6585$, [O\,$\rm II$] and [O\,$\rm III$] doublets, [S\,$\rm II$]$ \, 6718$  (see Fig. \ref{fig:spelines}). Fluxes were measured using the \texttt{slinefit}\footnote{\url{https://github.com/cschreib/slinefit}} code \citep{Schreiber2018} which also allows the correction of the Balmer-line fluxes due to the underlying stellar absorption. All the measured fluxes are reported in Tab. \ref{tab:spelines}.
Using the Balmer decrement and assuming the theoretical value for the (H$\alpha$/H$\beta$) ratio in the absence of dust, for case B recombination at a gas temperature of $10^4$ K \citep{Osterbrock1989}, we derived an intrinsic A$_{\rm V} = 2.8 \pm 0.3$.

\begin{figure}[t]
\centering
\includegraphics[height=8cm]{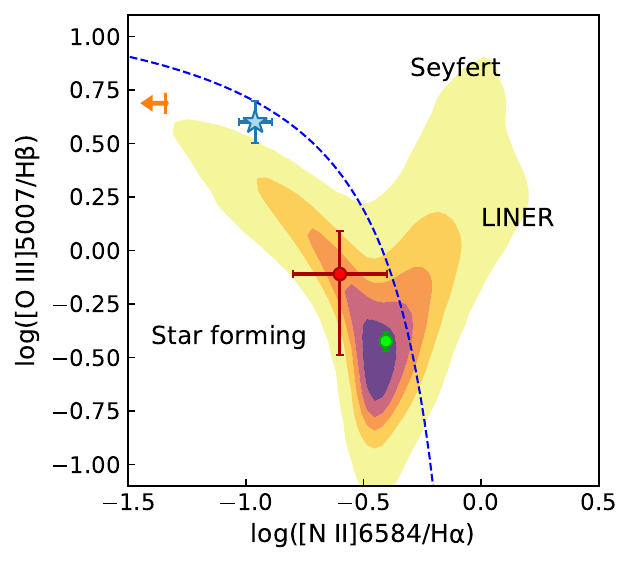}
\caption{Baldwin-Phillips-Terlevich empirical optical emission-line diagnostic diagram using the optical line ratios [N\,$\rm II]/H\alpha$ and [O\,$\rm III]/H\beta$. Shade areas represent the distribution of galaxies with both $u^\prime$ and $r^\prime$ meaningful Petrosian magnitudes in the SDSS catalogue. The dashed line, computed by \cite{Kauffmann2003}, delimits the star-forming galaxies region. The FRB\,20240114A host galaxy is marked by a filled star. The orange arrow marks the [N\,$\rm II]/H\alpha$ upper limit for the FRB\,20121102A host \citep{Tendulkar21}; FRB\,20181030A \citep{Bhardwaj21_20181030Ahost} and FRB\,20201124A \citep{Bruni2024Nat} hosts are marked in red and green, respectively.
}
\label{fig:BPT_NII}
\end{figure}

\begin{figure}[t]
\centering
\includegraphics[height=8cm]{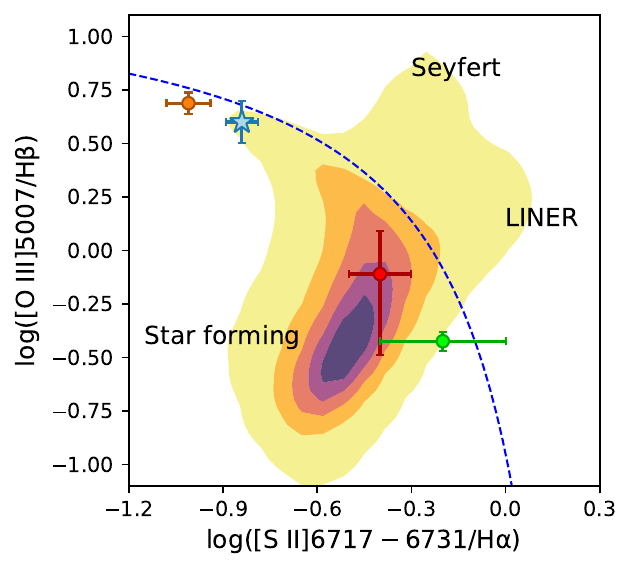}
\caption{Baldwin-Phillips-Terlevich diagram of diagram showing the ratios [S\,$\rm II]/H\alpha$ and [O\,$\rm III]/H\beta$. Background galaxies selection and FRB host markers as in Fig. \ref{fig:BPT_NII}. S\,II data for FRB\,20201124A are from \cite{Bhandari22}. The maximum starburst line, computed by \cite{Kewley2006}, is shown.}
\label{fig:BPT_SII}
\end{figure}

In addition to the [N\,$\rm II]/H\alpha$ versus [O\,$\rm III]/H\beta$ BPT diagram of Fig. \ref{fig:BPT_NII}, the location of the source in the [S\,$\rm II]/H\alpha$ vs [O\,$\rm III]/H\beta$ diagram is shown in Fig. \ref{fig:BPT_SII}. Public data for three additional PRS hosts are shown. Please note that FRB\,20190417A host galaxy falls in the star-forming region too \citep{Ibik2024}, though the line fluxes are not present in the literature. Moreover, \cite{niu22} provides only the $\rm H\alpha$ line flux for FRB\,20190520B.

\subsection{LBC photometry}
Photometric follow-up was obtained in optical with the LBC camera mounted on LBT.
Data were obtained in the $ g^\prime r^\prime i^\prime z^\prime$ bands on November 9, 2024, and additionally in the $ r^\prime i^\prime$ bands on the following night under not-excellent sky conditions.
LBC imaging data were processed using the data reduction pipeline developed at INAF - Osservatorio Astronomico di Roma \citep{Fontana2014a} which includes bias subtraction and flat-fielding, bad pixel and cosmic ray masking, astrometric calibration, and coaddition. Only the images showing FWHM below 1\farcs2 were selected and used, summing up to a total net exposure of 360 s in $g^\prime z^\prime$-bands, and 1100 and 1200 in $r^\prime i^\prime$-bands, respectively. Field stars from the GAIA DR3 catalogue \citep{Gaia2023j} were used for astrometry calibration. The resulting images average astrometric precision is 0\farcs05.

All LBC data were analysed by first performing aperture photometry using \texttt{DAOPHOT} and \texttt{APPHOT} under \texttt{PyRAF/IRAF} \citep{Tody1993} which was used for the photometric calibration against the \textit{Pan-STARRS} survey  \citep{Panstarrs2016,Panstarrsdatabase2020} for $r^\prime i^\prime z^\prime$ bands and \textit{SDSS DR16} \citep{sdssdr16} for $g^\prime$-band\footnote{The $g^\prime$ from Pan-STARRS does not match exactly the SLOAN filter, though the magnitude offset is smaller than the error.}, after careful selection of a set of isolated field stars.
For the host-galaxy we applied Petrosian-like elliptical photometry using \texttt{Source-Extractor} \citep{Bertinsex1996}. It is well detected in all bands and we measure the following AB magnitudes:
22.38$\pm$0.05, 
21.86$\pm$0.04,   
21.79$\pm$0.04, 
21.76$\pm$0.10, 
respectively in the $g^\prime r^\prime i^\prime z^\prime$-filters and not corrected for the foreground Galactic extinction. These values are consistent, within errors, with those reported by the \textit{Pan-STARRS} survey.
As a cross-check, we downloaded the \textit{SDSS} images and independently measured consistent magnitudes, within errors.

\subsection{SED fitting}
\begin{figure}[t]
    \centering
    \includegraphics[width=0.95\linewidth,angle=0]{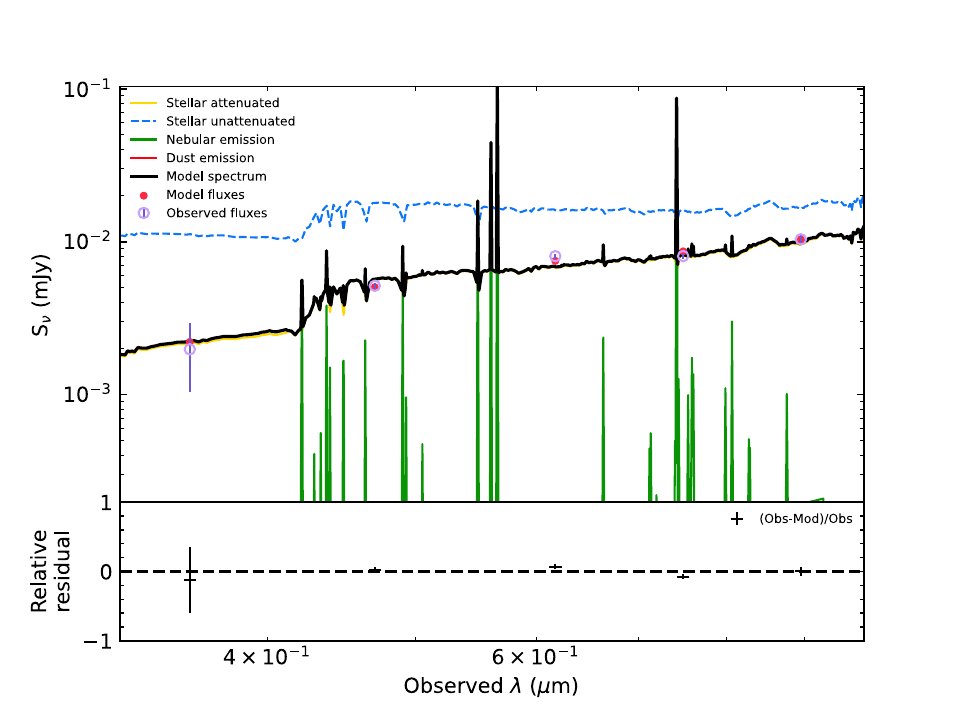}
\caption{SED fitting of the SDSS $u^\prime$ and LBC $g^\prime\,r^\prime\,i^\prime\,z^\prime$ photometry of the host galaxy with {\tt Cigale}. The best model (solid line) is shown along with the model predicted magnitudes (filled bullets). Photometric measurements are marked by violet empty circles.
}
\label{fig:sedhost}
\end{figure}

After correcting our $u^\prime g^\prime r^\prime i^\prime z^\prime$ data-set for the Galactic foreground extinction of $\mathrm{E(B-V)}=0.06$ mags \citep{sf2011}, we modelled the SED of the host galaxy using the code \texttt{Cigale}\footnote{\url{https://cigale.lam.fr/}} 
\citep{Boquien2019} with a redshift fixed to $z=0.13$.
The $u^\prime=23.0\pm0.4$ SDSS detection (model magnitude) was added to our dataset.
At first, we modelled the SED with a free $\mathrm{E(B-V)_{star}}$, and afterwards we fixed its value and used a derived stellar dust attenuation derived from the one measured with the Balmer decrement, $\mathrm{E(B-V)_{star}}=0.4 \times \mathrm{E(B-V)_{gas}}$ \citep{Calzetti1997, Asari2007}. 
Both modelling gave a similar and well constrained mass of $M_\star = 10^{8.1 \pm 0.2}\,{\it M}_\odot$.
The obtained SFR, attenuation and age are consistent between the two models, but loosely constrained by their large errors.
Fixing the metallicity to the value found via spectroscopy did not improve the fit.
Figure \ref{fig:sedhost} shows the best-fit model we obtained  and fixed $E(B-V)$. It provides a $\chi^2/{\rm dof}=0.25$ 
and a star-formation rate $\rm{SFR} < 1 \, {\it M}_{\odot}\,{\rm yr}^{-1}$. While the derived stellar mass is reliable, the SFR is degenerate with age (0.3--1.2 Gyr). 
We found consistent results by modelling the SED with the \texttt{LePhare} code \citep{Arnouts1999a, Ilbert2006a}. The photometric data are best-fit ($\chi^2/N_{filt}=4.9/5$) by a galaxy with the same stellar mass found above. Also in this case SFR, age and dust-extinction are degenerate and unconstrained.

%


\begin{table}
\caption{Emission lines fluxes and uncertainties in units of $10^{-17}$ [$\rm erg~s^{-1}~cm^{-2}$]. Values corrected for Galactic extinction only.}

\centering
\begin{tabular}{lc}
\hline
Line &  Fluxes   \\
\hline

[O II] 3727 & $~4.8 \pm 0.4 $ \\

[O II] 3729 & $~6.4 \pm 0.4 $ \\

H$\beta$ & $~2.5 \pm 0.4 $  \\

[O III] 5007 & $11.2 \pm 0.6 $ \\

[O III] 4959 & $~3.8 \pm 0.4 $ \\

H$\alpha$ & $19.8 \pm 0.3$  \\

[N II] 6549 & $~0.6 \pm 0.1 $ \\

[N II] 6584 & $~2.2 \pm 0.3 $ \\

[S II] 6718 & $~2.9 \pm 0.3 $ \\
\hline
\end{tabular}
\label{tab:spelines}
\end{table}


\section{The nebular model: Calculations}
\label{calculations}

The PRS has been proposed to be generated by the synchrotron emission (e.g., \citealt{yang20}), thus, the peak frequency $\nu_{\rm peak}$ should correspond to the typical frequency of the synchrotron from the electrons with the minimum Lorentz factor, $\gamma_m$, or the typical frequency of the synchrotron self-absorption, $\nu_a$. In the former scenario, one has
\begin{align}
    \nu_m=\frac{\gamma_m^2eB}{2\pi m_ec}\sim\nu_{\rm peak}\lesssim0.65~{\rm GHz},
\end{align}
where $B$ is the magnetic field strength at the emission region. Therefore, we obtained the following constraints:
\begin{align}
    \left(\frac{\gamma_m}{10^3}\right)^2\left(\frac{B}{1~{\rm mG}}\right)\lesssim0.23.
\end{align}
For the latter scenario, the typical frequency of synchrotron self-absorption should be also below the uGMRT band, $\nu_a\lesssim0.65~{\rm GHz}$.
The properties of the PRS emission region could be studied via the observed constraint on $\nu_a$. 
Assuming that the electron in the nebula satisfies a power-law distribution with $dn_e/d\gamma=(p-1)(n_{e,0}/\gamma_m)(\gamma/\gamma_m)^{-p}$ for $\gamma\geqslant\gamma_m$, where $n_{e,0}$ is the total electron number density, then the optical depth of the synchrotron radiation is \citep[e.g.,][]{yan16}
\begin{align}
    \tau_{\nu}=\frac{e^2(p-1)n_{e,0}\gamma_m^{p-1}R}{4m_ec}\frac{1}{\nu_B}\left(\frac{\nu}{\nu_B}\right)^{-(p+4)/2}f_{\alpha}(p),  
\end{align}
where $R$ is the radius of the nebula, $\nu_B=eB/2\pi m_ec$ is the electron cyclotron frequency in a magnetic field $B$, $f_{\alpha}(p)\equiv3^{(p+1)/2}\Gamma[(3p+2)/12]\Gamma[(3p+22)/12]$. 
The self-absorption frequency $\nu_a$ is defined by $\tau_{\nu}=1$, leading to
\begin{align}
    \nu_a=\nu_B\left[\frac{\pi e(p-1)n_{e,0}\gamma_m^{p-1}R}{2B}f_{\alpha}(p)\right]^{2/(p+4)}\sim\nu_{\rm peak}\lesssim0.65~{\rm GHz}.
\end{align}
As pointed out above, the synchrotron heating process of FRBs in the nebula requires $\nu_a\lesssim0.65~{\rm GHz}$. Assuming $p\sim2$ for a typical particle's acceleration mechanism, one finally has the following constraints
\begin{align}
    \left(\frac{B}{1~{\rm mG}}\right)^{2/3}\left(\frac{n_{e,0}}{10^3~{\rm cm^{-3}}}\right)^{1/3}\left(\frac{R}{10^{16}~{\rm cm}}\right)^{1/3}\left(\frac{\gamma_m}{10^3}\right)^{1/3}\lesssim0.6.
\end{align}

 Since the PRS is likely powered by synchrotron radiation due to their non-thermal spectra at radio bands
\citep{yan16, yang20, yang22, Murase16, metzger17},
the PRS brightness will depend on the electron distribution and the parallel strength of local magnetic fields. It has been suggested that |RM| could be a good proxy of the PRS specific luminosity $L_\nu$, with a predicted scaling law of 
\cite{yang20, yang22},
\begin{align}
L_{\nu}&= \frac{64\pi^3}{27}\zeta_e\gamma_{\rm c}^2m_ec^2R^2\left|{\rm RM}\right|\nonumber \\
&\simeq5.7\times10^{28}~{\rm erg~s^{-1}~Hz^{-1}} \zeta_e\gamma_{\rm c}^2 \left(\frac{\left|{\rm RM}\right|}{10^4~{\rm rad~m^{-2}}}\right)\left(\frac{R}{10^{-2}~{\rm pc}}\right)^2, \label{lum} 
\end{align}
where $R$ is the radius of the region that contributes to the PRS and the RM, $\zeta_e$ is the electron fraction that generates synchrotron emission in the GHz band which can be approximately given by
$\zeta_e\sim\gamma_{\rm obs}n_e(\gamma_{\rm obs})/n_{e,0}$ with $\gamma_{\rm obs}\sim(2\pi m_ec\nu_{\rm obs}/eB)^{1/2}$ for the observed frequency $\nu_{\rm obs}\sim1~{\rm GHz}$, 
$\gamma_{\rm c}$ is the Lorentz factor defined by $\gamma_{\rm c}^2\equiv\int n_e(\gamma)d\gamma/\int[n_e(\gamma)/\gamma^2]d\gamma$ \citep[see the method of][]{Bruni2024Nat}, and $n_e(\gamma)$ is the differential distribution of electrons, and $n_{e,0}$ is the total electron number density. In particular, for an electron distribution with a thermal and a non-thermal component, $\gamma_{\rm c}$ would be the typical Lorentz factor separating the two components. Such a predicted scaling has been recently verified by the detection of several more PRSs \citep{Bruni2024Nat, Ibik2024}.

\end{appendix}

\end{document}